\documentclass[12pt]{article}
\usepackage{amsmath}
\usepackage{mathrsfs}
\usepackage{enumerate}
\usepackage{graphicx}
\begin{document}
\def\be{\begin{equation}}
\def\bea{\begin{eqnarray}}
\def\ee{\end{equation}}
\def\eea{\end{eqnarray}}
\def\d{\partial}
\def\eps{\varepsilon}
\def\la{\lambda}
\def\La{\Lambda}
\def\b{\bigskip}
\def\nn{\nonumber \\}
\def\p{\partial}
\def\t{\tilde}
\def\h{{1\over 2}}
\def\be{\begin{equation}}
\def\bea{\begin{eqnarray}}
\def\ee{\end{equation}}
\def\eea{\end{eqnarray}}
\def\b{\bigskip}
\def\u{\uparrow}
\def\AA{\mathscr{A}}
\def\DD{\mathscr{D}}
\def\FF{\mathscr{F}}
\def\LL{\mathscr{L}}
\def\CC{\mathscr{C}}
\def\MM{\mathscr{M}}
\newcommand{\comment}[2]{#2}

\begin{center}

\vspace{1cm}
{\Large\bf
Black branes with cosmological constant
}
\vspace{1.4cm}

{Rhucha Deshpande\footnote{rdeshpande@albany.edu} and Oleg Lunin\footnote{olunin@albany.edu} }\\

\vskip 0.5cm

{\em  Department of Physics,
University at Albany (SUNY),
Albany, NY 12222, USA
 }

\vskip 0.08cm

\vspace{.2cm}

\end{center}

{\baselineskip 11pt
\begin{abstract}
\noindent
We study neutral black branes with flat and curved worldvolumes in the presence of a negative cosmological constant. We reduce the equations governing the dynamics of such objects to one second--order ODE and perform various asymptotic expansions of the resulting equation. We also analyze regular geometries which have the same symmetries as the branes and interpolate between an empty interior and AdS asymptotics. We show that the dynamics of such spacetimes is governed by the Abel equation.   
\end{abstract}
}


\newpage

\setcounter{tocdepth}{2}
\tableofcontents


\section{Introduction}

Black holes and black branes are fascinating laboratories for exploring classical and quantum effects caused by strong gravitational fields, and they have led to many important discoveries over the last few decades. There are many similarities between black holes and branes at a technical level, but there are also significant physical differences. In particular, starting with a Ricci--flat black hole, one can easily promote it into a black string or a brane by adding an appropriate number of flat directions to the metric. While the resulting configuration solves the Einstein's equations, the final geometry might have classical instabilities \cite{GrLa}\footnote{The system becomes stable if the longitudinal directions along the brane are compactified on a sufficiently small torus \cite{GrLaComp}}, in contrast to the original black hole. The higher--dimensional black branes can be constructed in a similar fashion, and they also suffer from instabilities \cite{GrLa}. 

The situation becomes even more interesting in the presence of fluxes. Starting with a charged black hole and adding flat directions, one arrives at a geometry that no longer solves equations of motion. In special cases, the cure for this problem is well known: the extra directions must be warped, and implementation of this procedure led to discovery of black $p$--branes \cite{HorStr}, which played an important role in stimulating the second string revolution in the 
1990s\footnote{In particular, it was crucial that the black $p$--branes turned out to be alternative descriptions of D--branes \cite{PolchEquiv}, which had been discovered earlier \cite{PolchBrane}.}. Furthermore, one can start with a Ricci--flat black hole, add flat directions, and perform various dualities to construct gravitational backgrounds produced by charged branes \cite{Sen}. This leads to an interesting question: can one start with a black hole that is already supported by fluxes and add extra dimensions to it? The general answer to this question is not known. Perhaps the simplest setting for exploring this situation involves the setting where the role of flux is played by a cosmological constant: just like fluxes, it creates an obstacle to adding flat extra dimensions by making the curvature components in these directions proportional to the metric\footnote{Recall, that if one starts with a $d$--dimensional geometry supported by a $p$--form flux $F_{\mu_1\dots \mu_{p}}$ and tries to add a flat extra dimension $x$, then the obstruction comes from 
the $(xx)$ component of the stress--energy tensor, $T_{xx}\propto g_{xx}F_{\mu_1\dots \mu_{p}}F^{\mu_1\dots \mu_{p}}$, which has some similarities with the cosmological constant term.}. Apart from mimicking the effect of fluxes,  
the systems with cosmological constant $\Lambda$ are interesting in their own right, with applications ranging from observational cosmology to string theory, AdS/CFT correspondence, and $f(R)$ gravity. The desire to get insights into promoting black holes with fluxes into black branes, as well as numerous applications of backgrounds with $\Lambda$, motivate the general study of black branes in the presence of cosmological constant carried out in this article. 

Interestingly, incorporation of cosmological constant in black hole geometries is rather straightforward. The Schwarzschild--AdS solution was constructed in early days of general relativity, and incorporation of the cosmological constant in rotating geometries was accomplished by Carter \cite{CarterLam} soon after Kerr's original discovery \cite{Kerr}. The extension of Ricci-flat rotating rotating black holes in higher dimensions \cite{MPerry} to non--zero values of $\Lambda$ took longer, but the result is remarkably natural and simple \cite{PopeAdS}. Furthermore, dynamical properties of black holes with and without cosmological constant are very similar: these backgrounds lead to separable and fully solvable equations of motion for scalar \cite{CarterLam,Carter,Kubiz}, vector \cite{Maxwell}, and tensor \cite{OLforms} fields. These similarities provide an additional motivation for finding geometries of black branes with a cosmological constant and for comparing their properties with known solutions with $\Lambda=0$. 

The study of extended black objects in the presence of cosmological constant has a long history, which began with exploration of black strings with AdS$_5$ asymptotics \cite{CopHor}. This analysis was then extended to black strings in arbitrary dimensions in \cite{MannRad}, and stability of these objects was studied in \cite{BlcStrLaStab}. Unfortunately, unlike the metrics for black branes with $\La=0$ and AdS black holes, the solutions describing neutral black branes with cosmological constant are not known analytically, so articles \cite{CopHor,MannRad,BlcStrLaStab} relied on numerical results. The goal of our paper is to provide some analytical insights that complement the numerical studies presented in \cite{CopHor,MannRad}. In particular, these articles argued for existence of very interesting {\it regular} geometries which preserve 
${\cal B}=R_t\times S^{d-p-2}\times R^{p}$ symmetry and asymptote to AdS$_d$.\footnote{Strictly speaking, articles \cite{CopHor,MannRad} analyzed only the $p=1$ case.}.  In the AdS/CFT setting, such solutions would describe a gravity dual of a vacuum in a field theory defined on ${\cal B}$. While the bulk geometries for $p=0$ and $p=d-3$ are well known (they describe the AdS space in global coordinates and in the Poincare patch), solutions for other values of $p$ have been explored only numerically \cite{CopHor,MannRad}. In this article we introduce a new parameterization of such spaces and demonstrate that the improved variables satisfy a relatively simple Abel equation\footnote{This equation covers a large class of nonlinear first--order ODEs.}, whose solutions have five fixed points. Although we were not able to find the analytical solution of this Abel equation, we demonstrate that for every $p$ and $d$, there is a unique solution regular in the interior, and at large values of $r$ it is necessarily driven to a fixed point of the Abel equation describing the geometry with AdS asymptotics. Therefore we provide an analytical proof of existence and uniqueness of regular geometries with symmetries ${\cal B}$, which were explored numerically in \cite{CopHor,MannRad}. We also analyze various properties or regular geometries and black branes with symmetries ${\cal B}$ in the presence of a negative cosmological constant. 

\bigskip

This paper has the following organization. In section \ref{SecFlatBranes} we analyze the solutions with the symmetries of flat $p$--branes (${\cal B}=R_t\times S^{d-p-2}\times R^{p}$) in the presence of a negative cosmological constant. After writing equations of motion and introducing convenient coordinates, in section \ref{Sec5d} we focus on a simple example of $(p,d)=(1,5)$, which was studied in \cite{CopHor} using numerical techniques. We demonstrate that in this case the regular space with AdS asymptotics is described by solutions of Abel equation which interpolate between two specific fixed points, and we show that once the regularity condition in the interior is imposed, the solution with AdS$_5$ asymptotics exists, and it is unique. The dynamics of a black string is governed by a second order nonlinear ODE, in contrast to a first--order Abel equation, and we show that once a regular expansion is imposed at the horizon, the solution necessarily approaches AdS$_5$. This again demonstrates existence and uniqueness of geometries describing black branes with arbitrary horizon size. We also find approximate expressions for the regular and black brane solutions in various regions. 

In section \ref{SecSubBrane} we repeat the same analysis for arbitrary values of $p$ and $d$. While the logic is exactly the same as in the $(p,d)=(1,5)$ example, the intermediate formulas are much more complicated, so we present only the final results. In the $p=1$ case, one recovers the black strings in arbitrary dimensions, which were studied numerically in \cite{MannRad}.

In section \ref{SecSubBlackHoles} we show that degenerate cases of our solutions reproduce some well-known geometries. 

In section \ref{SecCurvBrane} we study a different but related system. Motivated by some known geometries describing charged black branes in AdS space, we look at branes with AdS$_{p+1}$ rather than 
$R_t\times R^{p}$ worldvolume and impose the symmetry ${\cal B}'=$AdS$_{p+1}\times S^{d-p-2}$ rather than ${\cal B}$. While we find some local structure for such solutions with AdS$_{d}$ asymptotics, none of them are well--defined globally due to naked singularities.

The implications of our work and some open questions are discussed in section \ref{SecDiscus}. 

\section{Black branes with flat worldvolumes}
\label{SecFlatBranes}

Branes play very important role in string theory, and their dynamics has been extensively studied via worldsheet, 
DBI, and supergravity techniques. From the string theory perspective, neutral black branes arise from brane--antibrane systems, where tachyon condensation makes the worldsheet and DBI analyses challenging. In this article we will focus on the gravitational picture, where neutral branes are described by metrics without fluxes. In this section we will assume translational and rotational invariance in spacial directions longitudinal to the brane, and in section \ref{SecCurvBrane} we will extend the analysis to some branes with curved worldvolumes. 

While flat branes have been extensively studied in type II supergravities, very little is known about solutions in the presence of a cosmological constant \cite{CopHor,MannRad}, and the existing results heavily rely on numerics. Although an explicit analytical solution for such branes seems to be beyond reach due to nonlinearities of the underlying equations, in this section we will use some analytical approximations to extract various properties of flat branes with cosmological constant.

\subsection{Equations of motion}

Let us consider a stack of uniform $p$--branes in $D$ dimensions. Imposing the translational and rotational symmetries in the longitudinal spacial directions, as well as rotational invariance in transverse directions, we arrive at the ansatz
\bea\label{TheAnsatz}
ds^2=-f_1 dt^2+f_2(dx_1^2+\dots dx_p^2)+\frac{1}{f_3} dr^2+r^2 d\Omega_q^2
\eea
Here $(f_1,f_2,f_3)$ are functions of the radial coordinate $r$. This metric leads to a diagonal Ricci tensor, and its components are given by
\bea\label{RicciFlat}
R_{ab}&=&h_{ab}{f_3}\left[(q-1)(\frac{1}{f_3}-1)-\frac{rf_3'}{2f_3}-\frac{rf_1'}{2f_1}-\frac{prf_2'}{2f_2}\right],\nn
R_{ij}&=&\delta_{ij}\frac{f_2f_3}{4}\left[-\frac{(p-2) f_2'^2}{f_2^2}  - \frac{f_2' f_3'}{f_2f_3}   + 
\frac{f_2'}{f_2} (-\frac{2 q}{r}- \frac{f'_1}{f_1}) - \frac{2 f_2''}{f_2}\right],\\
R_{rr}&=&\frac{1}{4}\left[ \frac{f_1'^2}{f_1^2} - \frac{2 f_1''}{f_1} - \frac{f_3'}{f_3}(\frac{2 q}{r} + \frac{f_1'}{f_1} 
+ \frac{p f_2'}{f_2}) + \frac{p (f_2'^2 - 2 f_2 f_2'')}{f_2^2}\right],\nn
R_{tt}&=& \frac{f_1f_3}{4} \left[\frac{f_1' f_3'}{f_1f_3} - 
   \frac{f_1'^2}{f_1^2} + \frac{f_1'}{f_1}
 (\frac{2 q}{r} + \frac{p f_2'}{f_2}) + 
     \frac{2 f_1''}{f_1} \right].\nonumber
\eea
Here we used indices $(i,j)$ to label the spacial directions along the branes and indices $(a,b)$ to label the coordinates on the sphere. We also defined the metric $h_{ab}$ on the unit $q$--dimensional sphere: 
\bea
d\Omega_q^2=h_{ab}dy^ady^b.
\eea
In the absence of cosmological constant, the Einstein's equations read $R_{\mu\nu}=0$, and their solutions are briefly discussed in the Appendix \ref{AppFlat}. The physically-interesting geometries describe the 
well-known black branes \cite{HorStr}
\bea
ds^2=-fdt^2+\frac{dr^2}{f}+(dx_1^2+\dots dx_p^2)+r^2 d\Omega_q^2,\quad
f=1-\frac{M}{r^{q-1}}\,.
\eea
In this article we focus on black branes with a negative cosmological constant $\Lambda$, so we impose the Einstein's equations
\bea\label{EinstLam}
R_{\mu\nu}=\frac{2}{d-2}\La g_{\mu\nu}\,.
\eea
To simplify subsequent formulas, it is convenient to introduce a rescaled cosmological constant 
${\bar\La}$ and a length scale $L$ associated with it as
\bea\label{EinstLamDef}
{\bar\La}=\frac{2}{d-2}\La=-\frac{d-1}{L^2}
\eea
We will use ${\bar\La}$ and $L$ interchangeably.

Let us outline the reduction of the system (\ref{RicciFlat}), (\ref{EinstLam}) to a differential equation for a single function. The spherical components of equations (\ref{EinstLam}) suggest a convenient change of variables from $f_2$ to $g$: 
\bea\label{k2def}
f_2(r) = \left[\frac{g(r)}{f_1(r)f_3(r)}\right]^{\frac{1}{p}}
\eea
After this substitution, the spherical components of (\ref{EinstLam}) lead to an algebraic expression for $f_3(r)$ in terms of the new function $g(r)$:
\bea\label{f3final}
f_3(r)=\frac{2(-1+q+{\bar\La} r^2)g(r)}{2(-1+q)g(r) + r g'(r)}
\eea
The time component of (\ref{EinstLam}) gives a first order equation for function $g$, which can be easily integrated leading to an explicit, although complicated, expression for $g$. This leaves only one unknown function $f_1$ which satisfies one non--linear differential equation of third order, but interestingly, the order can be reduced by making a substitution:
\bea\label{f1AsH}
f_1(r) = (1- {\bar\La} r^2) \exp\left[{\int \frac{dr}{r h(r)}}\right]
\eea
The resulting equation for function $h(r)$ is rather complicated, so in the remaining part of this section we will focus on a slightly different, although less intuitive, variables, but here we stress that the expressions for all metric components in terms of $h$ are very simple. For example, in the $q=2$ case, they are given by (\ref{k2def}), (\ref{f1AsH}), and 
\bea
g(r)=\frac{h^2}{r^2}\exp\left[\int \frac{8{\bar\La} r^2 h dr}{(1- {\bar\La} r^2)^2}\right],\quad
f_3=\frac{(1-{\bar\La} r^2)^3 h}{r[4\bar\La r h^2+(1-{\bar\La} r^2)^2 h']}\,.
\eea

In the remaining part of this section, we introduce a slight modification of the very intuitive ansatz (\ref{k2def}), (\ref{f1AsH}) and study the resulting equations in various limits. To conclude this subsection, we comment on the asymptotic form of the metric. For large values of $r$, the metric should approach AdS space, so in the leading approximation the geometry must have the form
\bea\label{AdSapprox}
ds^2 = -[\frac{r^2}{L^2} +a+..] dt^2+ \frac{dr^2}{\left[\frac{r^2}{L^2}+b+..\right]} + 
[\frac{r^2}{L^2}+c+..](dy_1^2+...+dy_p^2)+r^2 d\Omega_q^2
\eea
Direct substitution into the Einstein's equations (\ref{EinstLam}) gives the values of constants $(a,b,c)$:
\bea\label{AdScoeff}
a =c= \frac{q-1}{q+p-1},\quad
b = \frac{(q-1) (q + 2 p)}{(q+p-\frac{1}{2})^2 -\frac{1}{4}}.
\eea
In the leading order the metric (\ref{AdSapprox}) describes a space of constant curvature,
\bea
R_{\mu\nu\rho\lambda} =-\frac{1}{L^2} (g_{\mu\rho}g_{\nu\lambda} - g_{\mu\lambda}g_{\nu\rho}),
\eea
but these conditions are violated already by the $(a,b,c)$ terms unless $p=0$ or $q=1$. These special cases describe AdS space in global coordinates and on the Poincare patch\footnote{The Poincare patch can also be described by setting $q=0$ in (\ref{AdSapprox}), but since the sphere disappears for this value of $q$, one encounters a freedom in re-defining the radial coordinate. To avoid this freedom, we go to ``$q=0$ limit'' in (\ref{AdSapprox}) by setting $q=1$. The same trick will be used in section \ref{SecSubBlackHoles}.}.

\subsection{Example: black string in five dimensions}
\label{Sec5d}

The metric components in (\ref{TheAnsatz}) have the same qualitative behavior for all physically interesting values of $p$ and $q$, but the explicit dependence on these parameters is somewhat complicated. To illustrate the general procedure for solving the Einstein equations (\ref{EinstLam}), we begin with a specific case of a black string in five dimensions, 
$(p,q)=(1,2)$,
\bea\label{TheAnsatzStr}
ds^2=-f_1 dt^2+f_2 dx_1^2+\frac{1}{f_3} dr^2+r^2 d\Omega_2^2\,,
\eea
and a straightforward extension to the general case with more complicated formulas will be presented in section \ref{SecSubBrane}.

An extensive numerical study of black strings (\ref{TheAnsatzStr}) with cosmological constant was performed in \cite{CopHor}, so here we will focus on analytical aspects of the problem, which were not covered in that work. Specifically, we will analyze the case of two well-separated scales by assuming that $r_h\ll L$, where $r_h$ is the horizon radius. Then it is natural to look at two overlapping regions: $r\gg r_h$ and $r\ll L$.
First we notice that in the leading orders in $r_h/r$, the solution (\ref{AdSapprox})--(\ref{AdScoeff}) has 
$g_{tt}=g_{yy}$. This observation, along with analysis of data coming from numerical integration, 
suggests that 
\bea\label{temp19}
f_1(r)=f(r)+\frac{r_h}{r}{\tilde f}_1(r),\quad f_2(r)=f(r)+\frac{r_h}{r}{\tilde f}_2(r),
\eea
where function $f(r)$ does not depend on the horizon radius $r_h$. 

In section \ref{SecSubAbel} we will ignore the $(r_h/r)$ corrections in (\ref{temp19}) and reduce the dynamics of this problem to one first--order ODE. The final result is relatively simple, and it belongs to a well-known class of Abel equations. Although we motivate the Abel equation as an approximation for the $(r_h/r)$ limit, this equation becomes exact if the black string has zero horizon radius. The resulting geometry is regular everywhere, and an existence of such space (\ref{TheAnsatzStr}) interpolating between empty interior and AdS asymptotics was first demonstrated numerically 
in \cite{CopHor}. By showing that this spacetime is governed by the Abel equation, we provide an analytical handle for studying its properties. 

In section \ref{SecSubBlcStr} we will go beyond the small $(r_h/r)$ limit and reduce the exact problem to one nonlinear second--order ODE. We will then study various asymptotic expansions of this equation. 

\subsubsection{Regular geometry and the Abel equation}
\label{SecSubAbel}

As we have discussed above, in the small $(r_h/r)$ limit, functions $f_1$ and $f_2$ are approximately equal, so we begin with studying the geometry (\ref{TheAnsatzStr}) with 
\bea\label{temp19a}
f_2=f_1=f.
\eea
Function $f(r)$ does not depend on the horizon radius $r_h$, and it provides the {\it exact} description of the globally regular space with AdS asymptotics discovered in \cite{CopHor}. Substituting (\ref{TheAnsatzStr}) and (\ref{temp19a}) into the Einstein's equations (\ref{EinstLam}) with (\ref{RicciFlat}), we find a closed-form second-order equation for function 
$f$. Reducing the order of that equation by a substitution
\bea
f(r)=\exp\left[\int\frac{h(r)}{r}dr\right],
\eea
we arrive at the Abel equation of the first kind \cite{Abel} for function $h$:
\bea\label{AbelEqn}
4 (L^2 - 12 r^2) h + 
  8 (L^2 + 3 r^2) h^2 + (L^2 + 4 r^2) h^3 + 
  4 r (L^2 + 6 r^2) h'=32r^2.
\eea
The solution of this equation, $h(r)$, determines all metric components in the $r\gg r_h$ approximation through the relations
\bea\label{f123asH}
f_1(r)=f_2(r)=\exp\left[\int\frac{h(r)}{r}dr\right],\quad 
f_3(r) = \frac{4 (L^2 + 6 r^2)}{L^2 (4 + 8 h + h^2)}\,.
\eea
The system (\ref{AbelEqn})--(\ref{f123asH}) solves two problems: it provides an approximate description of the five--dimensional string (\ref{TheAnsatz}) for $r\gg r_h$ and describes the {\it exact} regular geometry interpolating between a patch of flat space near $r=0$ and AdS$_5$ at infinity. The existence of such regular $r_h=0$ limit of a black string was demonstrated in \cite{CopHor} using numerical study of complicated ODEs, but as we now see, this geometry is governed by a relatively simple system (\ref{AbelEqn})--(\ref{f123asH}). Let us now analyze the properties of Abel equation (\ref{AbelEqn}).

While many Abel equations admit analytic solutions \cite{Zaycev}, the special case (\ref{AbelEqn}) does not seem to fall into these solvable categories. To analyze the relevant configurations, it is useful to begin with making some general observations.
\begin{enumerate}[(a)]
\item Equation (\ref{AbelEqn}) admits exact solutions
\bea\label{AbelConst}
h(r)=2(\sqrt{3}-2),\qquad h(r)= -2(2+ \sqrt{3}).
\eea
\label{PageLabelAE}
\item At large values of $r$, function $h$ either diverges or approaches one of three constants:
\bea\label{AttractorInf}
r\gg L:&&h\rightarrow \quad -\infty,\quad -2(2+ \sqrt{3}),\quad 2(\sqrt{3}-2), \quad 2, \quad  \infty.
\eea
The AdS asymptotics in (\ref{f123asH}) correspond to the fixed point with $h=2$. 
\item At small values of $r$, function $h$ either diverges or approaches one of three constants:
\bea\label{AttractorZero}
r\ll L:&&h\rightarrow \quad -\infty,\quad -2(2+ \sqrt{3}),\quad 2(\sqrt{3}-2), \quad 0, \quad  \infty.
\eea
Equations (\ref{TheAnsatzStr}) and (\ref{f123asH}) describe regular geometry only if $h$ approaches zero at small values of $r$. 
\item Between small and large values of $r$, function $h$ interpolates between fixed points from sets (\ref{AttractorInf}) and (\ref{AttractorZero}) without crossing the constant solutions (\ref{AbelConst}). The physically interesting case corresponds to interpolation between $h=0$ and $h=2$. As we will see below, starting with a regular solution with $h=0$ at small values of $r$, one is always driven to $h=2$ fixed point at infinity.
\item A shift of $h$ by one of the constant solutions (\ref{AbelConst}), e.g., $h=2(\sqrt{3}-2)+u$, converts (\ref{AbelEqn}) into a homogeneous Abel equation for $u$. The coefficients in front of various powers of $h$ remain linear in $r^2$, but numerical factors become irrational. We will not use the equation for $u$ and focus on (\ref{AbelEqn}) instead.
\end{enumerate}
It is instructive to analyze the Abel equation (\ref{AbelEqn}) in the regions $r\gg L$ and $r\ll L$ beyond the leading terms (\ref{AttractorInf}), (\ref{AttractorZero}). 

The asymptotic form (\ref{AdSapprox}) for $f_1$ suggests that at large values of $r$ one should pick the $h=2$ option in (\ref{AttractorInf}), then the expansion of function $h$ reads
\bea\label{hRexpand}
h(r)=2-x+\frac{29x^3}{72}+\alpha x^2(1-\frac{4x}{3})+
\frac{x^2}{18}(3-4x)\ln x
+o(x^3),\quad x\equiv \frac{L^2}{r^2}
\eea
The solution is fully specified by the value of one parameter $\alpha$, and one can view (\ref{hRexpand}) as a double expansion in $x$ and in $\alpha$. The coefficient in front of $\alpha^k$ has the form 
$a_k x^{2k}[1+o(1)]$, and interestingly, one can find an explicit form for all coefficients $a_k$. To do so, we  take a {\it formal limit} of small $x$ while keeping $\alpha x^2$ fixed and consider an expansion
\bea\label{hSerInfty}
h_\infty(r)=2+\sum_{k=1}^\infty a_k \alpha^k \frac{L^{4k}}{r^{4k}}.
\eea
This function must satisfy a truncated version of (\ref{AbelEqn}),
\bea\label{AbelEqnRinf}
- 12 h_\infty + 6 h_\infty^2 + h_\infty^3 + 6 r h_\infty'=8.
\eea
and integration of this equation gives
\bea\label{AbelSolnRinf}
r^8=\frac{\mu_\infty}{(h_\infty-2)^{2}}
(h_\infty+4+2\sqrt{2})^{1-\sqrt{3}}(h_\infty+4-2\sqrt{2})^{1+\sqrt{3}}.
\eea
One integration constant $\mu_\infty$ determines all coefficients in (\ref{hSerInfty}), and comparison of leading terms in (\ref{hRexpand}) and (\ref{AbelSolnRinf}) gives
\bea\label{MuInfty}
\mu_\infty=L^8 \frac{\alpha^2}{
(6+2\sqrt{2})^{1-\sqrt{3}}(6-2\sqrt{2})^{1+\sqrt{3}}}
\eea
Expressions (\ref{AbelSolnRinf})--(\ref{MuInfty}) give the exact solution of equation (\ref{AbelEqnRinf}) and cover all contributions of the form $(a_k \alpha^k x^{2k})$ in the expansion (\ref{hRexpand}). 

A similar analysis of the Abel equation (\ref{AbelEqn}) for $r\ll L$ is slightly more involved. Neglecting $r^2$ in comparison to $L^2$, as well as the right--hand side of (\ref{AbelEqn}), and keeping $h=h_0(r)$ fixed, we find an approximate equation
\bea\label{AbelEqnR0}
&&4h_0 + 8h_0^2 + h_0^3 + 4 r h_0'=0,
\eea
and its solution reads
\bea
\label{AbelSolnR0}
&&r=\frac{\mu_0}{h_0}
(h_0+4-2\sqrt{3})^{\frac{1}{2}+\frac{1}{\sqrt{3}}}(h_0+4+2\sqrt{3})^{\frac{1}{2}-\frac{1}{\sqrt{3}}}.
\eea
For non--zero integration constant $\mu_0$, the solution $h_0$ approaches infinity (an allowed option in (\ref{AttractorZero})) when $r$ goes to zero. To make the geometry regular in the $r_h\rightarrow 0$ limit, we have to ensure that $h$ approaches zero at small values of $r$. This leads to $\mu_0=0$ and $h_0(r)=0$, but approximation (\ref{AbelEqnR0}) is not valid for this solution since the right--hand side of (\ref{AbelEqn}) can't be ignored. To get a good limit, we write $h={\tilde h}_0/L^2$ in (\ref{AbelEqn}) and send $L$ to infinity while keeping ${\tilde h}_0$ fixed. This gives
\bea\label{AbelSolnR0p}
4 {\tilde h}_0 + 
  4 r {\tilde h}_0'=32r^2 \quad \Rightarrow \quad h_0=\frac{8r^2}{3L^2}+\frac{\nu_0}{L^2 r}
\eea
To summarize, the solution of the Abel equation at $r\ll L$ is given either by (\ref{AbelSolnR0}) with 
$\mu_0\ne 0$ or by (\ref{AbelSolnR0p}). Both branches can be glued to AdS asymptotics  at $r\gg r_h$, but the regular solution with $r_h=0$ corresponds to (\ref{AbelSolnR0p}) with 
$\nu_0=0$. Substitution of (\ref{AbelSolnR0p}) into (\ref{f123asH}) and application of $L\rightarrow\infty$ limit gives
\bea\label{f123asHflt}
f_1(r)=f_2(r)=\mbox{const},\quad 
f_3(r) = 1\,.
\eea
Let us summarize the properties of the regular geometry with $r_h=0$.
\begin{enumerate}[1.]
\item The system (\ref{AbelEqn})--(\ref{f123asH}) describes the unique regular asymptotically--AdS geometry (\ref{TheAnsatzStr}) once the appropriate boundary conditions are imposed.
\item This regular solution approaches AdS space at large values of $r$, but it deviates from the maximally symmetric space in the interior. It is interesting to compare the ansatz (\ref{TheAnsatzStr}) with the Poincare patch of AdS space (where $S^2$ would be replaced by $R^2$) and with two types of global coordinates (where $y$ would be either a part of a sphere of combine with time direction to form AdS$_2$). While $R_t\times R^3$, $R_t\times S^3$, and $AdS_2\times S^2$ slicings lead to AdS space, the ansatz  (\ref{TheAnsatzStr}) does not. One may hope that the $AdS_2\times S^2$ slicing can be used to accommodate a curved black string, and such configurations will be discussed in section \ref{SecCurvBrane}.
\item Some general properties of the Abel equation (\ref{AbelEqn}) are summarized in the items (a)--(e) on page \pageref{PageLabelAE}.
\item At $r\ll L$, the regular solution is approximated by (\ref{AbelSolnR0p}) with $\nu_0=0$. Any other value of $\nu_0$ or a non--zero value of $\mu_0$ in the approximation (\ref{AbelSolnR0}) lead to a naked singularity at $r=0$. 
\item At $r\gg L$, the expansion is given by (\ref{hRexpand}), and the leading terms in the series can be summed to give (\ref{AbelSolnRinf})--(\ref{MuInfty}). Numerical integration of 
the Abel equation (\ref{AbelEqn}) from the approximation (\ref{AbelSolnR0p}) to large values of $r$ gives $\alpha\simeq 0.15$ in (\ref{hRexpand}). 
\end{enumerate}
Although the existence of a regular solution of the form (\ref{TheAnsatzStr}) which interpolates between 
AdS space at large values of $r$ and an empty space in the interior was proposed in \cite{CopHor}, where very impressive numerical evidence for this conjecture was presented, our parameterization (\ref{f123asH}) leads to the first {\it analytical proof} of existence. Specifically, we have demonstrated that in the new parameterization (\ref{f123asH})  the Einstein's equations for a horizon--free metric (\ref{TheAnsatzStr}) are reduced to a non--linear Abel equation (\ref{AbelEqn}), whose solutions interpolate between the fixed points (\ref{AttractorInf}) and (\ref{AttractorZero}) without crossing the exact solutions (\ref{AbelConst}). Starting from the approximation $h_0=\frac{8r^2}{3L^2}$ near $r=0$, the {\it unique} option for a regular geometry\footnote{Recall  the discussion around equation (\ref{AbelSolnR0p}).}, the solution is necessarily driven to the $h=2$ fixed point in (\ref{AttractorInf}), guaranteeing the AdS asymptotics at large values of $r$. Then the unique asymptotic expansion is given by (\ref{hRexpand}), and it is fully specified by one parameter 
$\alpha$, which is determined numerically ($\alpha\simeq 0.15$). In addition to proving the existence of the regular solution, we have also analyzed its properties covered by items 1--5 above. 

\subsubsection{Asymptotic expansions for black strings}
\label{SecSubBlcStr}
Let us now discuss black strings with non--zero horizon radius. We will assume that $r_h\ll L$  and consider two overlapping regions: $r\gg r_h$, where the Abel equation (\ref{AbelEqn}) still holds with high accuracy and $r\ll L$. In the second region we will no longer use the $f_1=f_2$ approximation. 

Without loss of generality, we can still impose the relation (\ref{f123asH}) for $f_2$, but not for $f_1$. Then a minor modification of arguments presented in section \ref{SecSubAbel} leads to the expressions for the functions $(f_1,f_2,f_3)$,
\bea\label{f123SecOrd}
&&f_1(r)=\frac{1}{r^2f_3(r)f_2(r)}
\exp\left[\int \frac{2(L^2+4r^2)h' dr}{8 r^2 - (L^2 + 4 r^2) h}\right]
\nn
&&f_2(r)=\exp\left[\int\frac{h(r)}{r}dr\right],\quad 
f_3(r) = \frac{8 r^2 - (L^2 + 4 r^2) h}{L^2 r h'}\,,
\eea
and to a closed--form {\it second order} ODE for function $h$
\bea\label{SecOrdH}
&&\left[
     8 r^2 h + (L^2 + 4 r^2)h^2 + (L^2 + 4 r^2) h^3\right]+ 
  r \left[8 (L^2 + 3 r^2) + 3 (L^2 + 4 r^2) h\right] h'\nn
  &&\qquad - r (4 + h) \left[-8 r^2 + (L^2 + 4 r^2)h\right] \frac{h''}{h'}=64r^2.
\eea
Any function $h$ satisfying the Abel equation (\ref{AbelEqn}) solves (\ref{SecOrdH}) as well, and in the large $r$ limit, expressions (\ref{f123SecOrd}) indeed lead to $f_1=f_2$. The counterpart of the expansion (\ref{hRexpand}), 
\bea\label{hRexpSecOrd}
h=2-x+\alpha x^2+\beta x^3+\frac{x^2}{18}(3-4x)\ln x+o(x^3),\quad x\equiv \frac{L^2}{r^2},
\eea
now contains two integration constants $(\alpha,\beta)$, and for $r_h\ll L$ the parameter 
$\beta$ is very close to the value corresponding to (\ref{hRexpand}). 

As before, the $r\ll L$ limit splits into two cases:
\begin{enumerate}[a)]
\item Taking the large $L$ limit in (\ref{SecOrdH}) while keeping $h$ fixed, we arrive at a counterpart of equation (\ref{AbelEqnR0}):
\bea\label{SecOrdAppr}
     h_1^2 + h_1^3+ 
  r \left[8 + 3 h_1\right] h_1' - r (4 + h_1) h_1 \frac{h_1''}{h_1'}=0,
\eea
The solution reads
\bea\label{Alt_h1}
r=\frac{\mu_1}{h_1}
\left[h_1+\la_1(1-\frac{1}{2\nu_1})\right]^{\frac{1}{2}+\nu_1}\left[h_1+\la_1(1+\frac{1}{2\nu_1})\right]^{\frac{1}{2}-\nu_1}{\hskip -0.7cm},\qquad
\nu_1=\frac{\la_1}{2\sqrt{(\la_1-1)^2+3}}\,,
\eea
and as expected, it has two integration constants. The expression (\ref{AbelSolnR0}) is recovered by setting $(\mu_1,\la_1)=(\mu_0,4)$. As we will see below, the solution (\ref{Alt_h1}) with nonzero values of $\mu_1$ leads to a wrong asymptotic behavior near $r=r_h$, so it should be discarded, just like (\ref{AbelSolnR0}).
\item To obtain the counterpart of the limit (\ref{AbelSolnR0p}), we write $h={\tilde h}_1/L^2$ and keep ${\tilde h}_1$ fixed when $L$ goes to infinity in (\ref{SecOrdH}). This gives
\bea\label{h1asymp0}
  8r {\tilde h}_1'- 4r \left[-8 r^2 + {\tilde h}_1\right] \frac{{\tilde h}_1''}{{\tilde h}_1'}=64r^2.
\eea
The solution of this equation is a generalization of (\ref{AbelSolnR0p})
\bea\label{hRhFlat}
h=\frac{1}{L^2}\left[\frac{8r^3}{3(r-r_h)}+\frac{C}{r-r_h}\right],
\eea
and it leads to the metric components (\ref{f123SecOrd}):
\bea\label{f123flat}
f_1=C_1\left[1-\frac{r_h}{r}\right],\quad f_3=1-\frac{r_h}{r},\quad f_2=C_2\,.
\eea
Here $(C,C_1,C_2,r_h)$ are integration constants. Substituting (\ref{f123flat}) into (\ref{TheAnsatzStr}), we recover the metric of the black string with $\Lambda=0$ 
in asymptotically-flat space. 
\end{enumerate}
Although integration constant $C$ does not appear in the expressions (\ref{f123flat}) for the metric components\footnote{The choice of $C$ would indirectly affect the values $C_1$ and $C_2$ when equation (\ref{SecOrdH}) is integrated between $r_h$ and infinity and the results are substituted in (\ref{f123SecOrd}). However, in the in the region $r\ll r_h$, $(C,C_1,C_2)$ are free parameters.}, a generic value of this parameter invalidates the approximation (\ref{h1asymp0}) near $r=r_h$ since $h$ becomes large. To avoid this problem, one must choose
\bea\label{hRhFlatC}
C=-\frac{8r_h^3}{3}\,.
\eea
While we are mostly interested in the $r_h\ll L$ case, where (\ref{hRhFlat}) works for all $r\ll L$, it is also instructive to find a counterpart of (\ref{hRhFlat}) that solves the full equation (\ref{SecOrdH}) in the vicinity of the horizon for all values of $L$. The expansion in powers of 
$\rho=(r-r_h)$ reads
\bea\label{GenNearHorH}
h=\frac{8 r_h^2}{L^2+4r_h^2}+\frac{8 L^2r_h\rho}{(L^2+4r_h^2)^2}+\frac{8L^2(L^2-6r_h^2)\rho^2}{3(L^2+4r_h^2)^3}-
\frac{160L^4r_h\rho^3}{9(L^2+4r_h^2)^4}+\dots
\eea
Only the first three terms contribute in the small $(r_h/L)$ limit, and they reproduce the solution (\ref{hRhFlat}) with integration constant (\ref{hRhFlatC}). Substitution of the last expression into (\ref{f123SecOrd}) gives a near--horizon expansion of the metric component $f_3$:
\bea\label{GenNearHorF3}
f_3=\frac{L^2+4r_h^2}{L^2 r_h}\rho-\frac{L^2+2r_h^2}{L^2 r^2_h}\rho^2+
\frac{9L^4+62L^2 r_h^2+144 r_h^4}{9L^2 r^3_h(L^2+4r_h^2)}\rho^3+\dots
\eea
For small $(r_h/L)$, this expansion reduces to the geometric series from (\ref{f123flat}). 
Note that while for generic $(L,r_h)$, the expansions (\ref{GenNearHorH}) (\ref{GenNearHorF3})
are applicable only for 
$\rho\ll L, r_h$, the solution (\ref{hRhFlat}), (\ref{hRhFlatC}), (\ref{f123flat}) is valid in a much larger region $r\ll L$, as long as $r_h\ll L$.

To find the metric (\ref{TheAnsatzStr}) for generic values of $(r_h,L)$, one should start with the expansion (\ref{GenNearHorH}), which contains only one parameter $r_h$ and integrate the differential equation (\ref{SecOrdH}) from $r_h$ to infinity. This procedure leads to the unique regular black string with AdS asymptotics, but unfortunately it can be carried out only numerically. Since a detailed numerical study of the system, albeit in different variables, was presented in \cite{CopHor}, we will not discuss these simulations further, but we will focus on the $r_h\ll L$ case instead. 

\bigskip
\noindent
Let us summarize the analytical results for black strings with $r_h\ll L$.
\begin{enumerate}[1.]
\item The geometry of the black string (\ref{TheAnsatzStr}) has the coefficients given by 
(\ref{f123SecOrd}), where function $h$ satisfies the second order ODE (\ref{SecOrdH}). 
\item In the $r\ll L$ region, the unique solution leading to a regular horizon is given by equations  (\ref{hRhFlat}) and (\ref{hRhFlatC}):
\bea\label{hRhFlatFin}
h=\frac{8[r^3-r_h^3]}{3L^2(r-r_h)}\,.
\eea
The resulting metric components are (\ref{f123flat}). The alternative approximation, (\ref{Alt_h1}) with non--zero value of $\mu_1$, does not reproduce the correct asymptotic expansion (\ref{GenNearHorH}) near a regular horizon, so it should be discarded.
\item By solving equation (\ref{SecOrdH}) beyond the leading order (\ref{hRhFlatFin}), one can find an expansion in inverse powers of $L$ 
\bea\label{BeyondLead}
h&=&\frac{8[r^3-r_h^3]}{3L^2(r-r_h)}-\frac{8}{135L^4}[
48r^4+108 r^3 r_h+193 (r_h r)^2+118 r_h^3 r+13 r_h^4]\nn
&&-\frac{16r_h^3(r+r_h)}{9L^4(r-r_h)}\ln\frac{r}{r_h}+O(\frac{1}{L^{6}}).
\eea
This series has no free parameters, apart from $r_h$, and as expected, the near--horizon expansion of (\ref{BeyondLead}) reproduces (\ref{GenNearHorH}).
\item In the $r\gg r_h$ region, equation (\ref{SecOrdH}) is well approximated by the first order Abel equation (\ref{AbelEqn}), and we have already discussed the properties of the solutions in section \ref{SecSubAbel}. In particular, for $r\gg L$, the solution can be expanded as (\ref{hRexpand}), which is the special case of (\ref{hRexpSecOrd}) with 
\bea\label{betaAsAlpha}
\beta=\frac{29}{72}-\frac{4\alpha}{3}\,.
\eea
\item Numerical integration of equation (\ref{SecOrdH}) with initial conditions (\ref{GenNearHorH}) near $r=r_h$ gives the unique solution which asymptotes to (\ref{hRexpSecOrd}) with specific values of $\alpha$ and $\beta$ for every $r_h$. The relation (\ref{betaAsAlpha}) is indeed reproduced for $r_h\ll L$.
\end{enumerate}

\subsection{Black $p$--branes in $d$ dimensions}
\label{SecSubBrane}

The analysis presented in section \ref{Sec5d} can be easily repeated for a more general geometry (\ref{TheAnsatz}). All steps remain the same, but the intermediate formulas look more convoluted due to complicated dependences on $p$ and $q$, so we present only the final results. As in section \ref{Sec5d} we give separate summaries for the unique globally regular geometry corresponding to $r_h=0$ and for the black branes with nontrivial horizon. 

\bigskip
\noindent
The solution for the regular horizon--free geometry has $f_2=f_1$, and in this case we proceed as in section \ref{SecSubAbel}.
\begin{enumerate}[1.]
\item Imposing the counterparts of relations (\ref{f123asH}),
\bea\label{f123asHpq}
f_1(r)&=&f_2(r)=\exp\left[\int\frac{h(r)}{r}dr\right],\\
f_3(r)&=&\frac{4}{L^2}\frac{L^2 q(q-1) + s(s-1) r^2}{4q (q-1)+ 4 q (1 + p) h + p (1 + p) h^2}\,,\nonumber
\eea
and substituting the result into the Einstein's equations, we arrive at the generalization of equation (\ref{AbelEqn}):
\bea\label{AbelEqnPq}
&&4 q \left[L^2 (q-1)^2 -(3 + 2 p - q) s r^2\right] h 
+ 2 (1 +p) \left[2 L^2 (q-1) q -(p - 2 q) s r^2\right] h^2\nn
&&\quad + p (1 + p) 
\left[L^2 (q-1) + s r^2\right] h^3 + 4 r\left[L^2 (q-1) q +(p + q) s r^2\right] h'\\
&&=8r^2 q(q-1) s\,. \nonumber
\eea
This is again the Abel equation \cite{Abel}. To simplify (\ref{AbelEqnPq}) and subsequent formulas, we defined a convenient combination $s=p+q+1=d-1$. The system (\ref{f123asHpq})--(\ref{AbelEqnPq}) describes the unique regular asymptotically--AdS geometry (\ref{TheAnsatz}) once the appropriate boundary conditions are imposed.
\item This regular solution approaches AdS space at large values of $r$, but deviates from the maximally symmetric space in the interior. In contrast to the well--known Poincare patch with the explicit 
$R^{d-2,1}$ isometries and to the global AdS with $AdS_p\times S^{d-1-p}$ slicings, the $R^{p,1}\times S^q$ isometries of (\ref{TheAnsatz}) don't lead to a simple solution, but rather reduce the problem to the system (\ref{f123asHpq}) governed by the Abel equation (\ref{AbelEqnPq}). 
\item Let us summarize the properties of the Abel equation (\ref{AbelEqnPq}), which generalize items (a)--(e) on page \pageref{PageLabelAE}:
\begin{enumerate}[(a)]
\item Equation (\ref{AbelEqnPq}) admits exact solutions
\bea\label{AbelConstPq}
h(r)=h_+,\quad h(r)=h_-,\quad h_\pm=-\frac{2q}{p}\pm\sqrt{\frac{4q(p+q)}{p^2(p+1)}}.
\eea
\item At large values of $r$, function $h$ either diverges or approaches one of three constants:
\bea\label{AttractorInfPq}
r\gg L:&&h\rightarrow \quad -\infty,\quad h_-,\quad h_+, \quad 2, \quad  \infty.
\eea
The AdS asymptotics in (\ref{f123asHpq}) correspond to the fixed point with $h=2$. 
\item At small values of $r$, function $h$ either diverges or approaches one of three constants:
\bea\label{AttractorZeroPq}
r\ll L:&&h\rightarrow \quad -\infty,\quad h_-,\quad h_+, \quad 0, \quad  \infty.
\eea
Equations (\ref{TheAnsatz}) and (\ref{f123asHpq}) describe regular geometry only if $h$ approaches zero at small values of $r$. 
\item Between small and large values of $r$, function $h$ interpolates between fixed points from sets (\ref{AttractorInfPq}) and (\ref{AttractorZeroPq}) without crossing the constant solutions (\ref{AbelConstPq}). The physically interesting case corresponds to interpolation between $h=0$ and $h=2$.  
\item A shift of $h$ by one of the constant solutions (\ref{AbelConstPq}), e.g., $h=h_++u$, converts (\ref{AbelEqnPq}) into a homogeneous Abel equation for $u$. 
\end{enumerate}
\item At $r\ll L$, the Abel equation can be simplified by taking limits in two different ways, and as in section \ref{SecSubAbel}, only one of them leads to a regular solution. We will provide some details in items (a) and (b), and the final answer is given by (\ref{AbelSolnR0pPq}). 
\begin{enumerate}[(a)]
\item
The simplest $r\ll L$ limit of the Abel equation (\ref{AbelEqnPq}) is taken by sending $L$ to infinity while keeping $h$ fixed. This gives
\bea\label{AbelEqnR0pq}
&&4 q  (q-1) h 
+ 4 (1 +p)q h^2+ p (1 + p) 
h^3 + 4 r q h'=0
\eea
and its solution reads
\bea
\label{AbelSolnR0pq}
&&r^{q-1}=\frac{\mu_0}{h_0}
\left[h_0+\frac{q}{p}(2-\frac{1}{\nu})\right]^{\frac{1}{2}+\nu}
\left[h_0+\frac{q}{p}(2+\frac{1}{\nu})\right]^{\frac{1}{2}-\nu}\,.
\eea
Here $\mu_0$ is an integration constant and 
\bea
\nu=\frac{1}{2}\sqrt{\frac{q(p+1)}{p+q}}
\eea
For non--zero integration constant $\mu_0$, function $h_0$ given by (\ref{AbelSolnR0pq}) approaches infinity at small values of $r$, and this leads to a singular geometry. For the 
$\mu=0$ solution, the approximation (\ref{AbelEqnR0pq}) fails, so we need an alternative rescaling. 

\item
To get a good limit, we write $h={\tilde h}_0/L^2$ in (\ref{AbelEqnPq}) and send $L$ to infinity while keeping ${\tilde h}_0$ fixed, as we did in section \ref{SecSubAbel}. This gives a differential equation
\bea\label{AbelSolnR0pPqInt}
(q-1) {\tilde h}_0 + r {\tilde h}_0'=2(p+q+1)r^2
\eea
which has the unique solution regular at $r=0$:
\bea\label{AbelSolnR0pPq}
h=\frac{2(p+q+1)r^2}{L^2(q+1)}+\frac{\nu_0}{L^2 r^{q-1}},\quad \nu_0=0.
\eea
To summarize, the unique regular geometry is described by the solution of the Abel equation which at $r\ll L$ behaves as (\ref{AbelSolnR0pPq}). The corresponding metric components (\ref{f123asHpq}) are
\bea\label{f123asHfltPq}
f_1(r)=f_2(r)=\mbox{const},\quad 
f_3(r) = 1\,.
\eea
\end{enumerate}
\item At $r\gg L$, the generalization of the expansion (\ref{hRexpand}) is given by  
\bea\label{hRexpandPq}
h = 2 - \frac{2(q-1)}{s-2}x  +\frac{2  (q-1)^2 [(s-1)(s-4)-2p]}{(s-2)^2 (s-1)(s-4)} x^2+ \dots,\quad
x=\frac{L^2}{r^2}\,.
\eea
Note that the denominator in the $x^2$ term vanishes for $(p,q)=(1,2)$, invalidating the expansion (\ref{hRexpandPq}) beyond linear order in $x$ and requiring logarithmic terms in (\ref{hRexpand}). 
For larger even values of $s$, the denominator of the 
$x^{k}$ term in (\ref{hRexpandPq}) contains a product $(s-2)(s-4)..(s-2k)$, which requires an introduction of a leading logarithmic contribution $[x^{s/2}\ln x]$ and allows a free parameter in front of $x^{s/2}$, a counterpart of $\alpha$ in (\ref{hRexpand}). For odd values of $s$, all terms in the power series (\ref{hRexpandPq}) are well defined, but one can add another series that starts with $r^s$ and contains only odd powers of $r$.

As in (\ref{AbelSolnRinf}), one can sum the series (\ref{hRexpandPq}) in the limit is small $x$ with fixed 
$\alpha x^s$, and the result reads
\bea\label{AbelSolnRinfPq}
r^{2s}=\frac{\mu_\infty}{(h_\infty-2)^{2}}
(h_\infty-h_+)^{1+w}(h_\infty-h_-)^{1-w},\quad w=\sqrt{\frac{(p+1)(p+q)}{q}}
\eea
As expected, in the leading order this gives $h_\infty=2+\alpha x^{s/2}+\dots$, where 
\bea
\alpha=\left[\frac{\mu_\infty}{L^{2s}}
(2-h_+)^{1+w}(2-h_-)^{1-w}\right]^{1/2}
\eea
The unique regular solution has the specific value of $\alpha$ which can be determined by performing a numerical integration of the Abel equation (\ref{AbelEqnPq}) from the approximation (\ref{AbelSolnR0pPq}) near $r=0$ to large values of $r$.
\end{enumerate}
The analysis presented here {\it proves} the existence of unique regular solutions of the form (\ref{TheAnsatz}) that interpolate between the  AdS space at large values of $r$ and an empty space in the interior, extending the results of section \ref{SecSubAbel} to all values of $(p,q)$. Specifically, we have demonstrated that the dynamical problem for a horizon--free metric (\ref{TheAnsatz}) is reduced to a non--linear Abel equation (\ref{AbelEqnPq}), whose solutions interpolate between the fixed points (\ref{AttractorInfPq}) and (\ref{AttractorZeroPq}) without crossing the exact solutions (\ref{AbelConstPq}). Starting from the approximation (\ref{AbelSolnR0pPq}) near $r=0$, the {\it unique} option allowed by regularity, the solution is necessarily driven to the $h=2$ fixed point in (\ref{AttractorInfPq}), guaranteeing the AdS asymptotics at large values of $r$. In addition to proving the existence of the unique regular geometries for all values of $(p,q)$ in (\ref{TheAnsatz}), the items 1--5 above also give various asymptotic expansions of the resulting solutions. 

\bigskip
\noindent
Let us now summarize the analytical results for black branes with nonzero $r_h$. These statements have been derived using the logic outlined in section \ref{SecSubBlcStr}.
\begin{enumerate}[1.]
\item The geometry of black branes (\ref{TheAnsatz}) has the coefficients given by 
the counterpart of (\ref{f123SecOrd}), 
\bea\label{f123SecOrdPq}
&&f_1(r)=\frac{1}{r^{2(q-1)}f_3(r)[f_2(r)]^p}
\exp\left[\int \frac{2[(q-1)L^2+sr^2]h' dr}{2s r^2 - [(q-1)L^2+sr^2] h}\right]
\nn
&&f_2(r)=\exp\left[\int\frac{h(r)}{r}dr\right],\quad 
f_3(r) = \frac{2s r^2 - [(q-1)L^2 + s r^2] h}{L^2 r h'}\,,
\eea
where function $h$ satisfies a second order ODE, the counterpart of (\ref{SecOrdH}):
\bea\label{SecOrdHpq}
&&
    2[qq_-(q-2)L^2+ s(p-q)^2r^2] h + p [L^2 q_-(2q-3) + (2q-2- p) s r^2]h^2\nn
    &&\qquad + \frac{p(p+1)}{2}[q_- L^2 + s r^2] h^3+ 
  r \left[4qq_-L^2 + 2(2q-p) r^2 + 3 (q_-L^2 + s r^2) h\right] h'\nn
  &&\qquad - r (2q + p h) \Big[-2s r^2 + [q_- L^2 + s r^2]h\Big] \frac{h''}{h'}=4q^2 s r^2.
\eea
To simplify the last expression we defined $q_-=q-1$.
\item In the $r\ll L$ region, 
the unique solution leading to a regular horizon is given by equation (\ref{hRhFlatPq}):
\bea\label{hRhFlatFinPq}
h=\frac{2s}{L^2(q+1)}\left[\frac{r^{q+1}-r^{q+1}_h}{r^{q-1}-r^{q-1}_h}\right].
\eea
It leads to the metric components (\ref{f123flatPq}). The alternative approximation, (\ref{Alt_h1Pq}) with non--zero value of $\mu_1$, does not reproduce the correct behavior near a regular horizon, so it should be discarded. Here are some details:
\begin{enumerate}[a)]
\item Taking the large $L$ limit in (\ref{SecOrdHpq}) while keeping $h$ fixed, we arrive at a generalization of equation (\ref{SecOrdAppr}):
\bea
\hskip -0.3cm
     2q(q-2)h_1+p(2q-3)h_1^2 + \frac{p(p+1)}{2}h_1^3+ 
  r \left[4q + 3 p h_1\right] h_1' - r (2q + p h_1) h_1 \frac{h_1''}{h_1'}=0.\nonumber
\eea
The solution reads
\bea\label{Alt_h1Pq}
r^{q-1}=\frac{\mu_1}{h_1}
\left[h_1+\la_1(1-\frac{1}{2\nu_1})\right]^{\frac{1}{2}+\nu_1}\left[h_1+\la_1(1+\frac{1}{2\nu_1})\right]^{\frac{1}{2}-\nu_1}{\hskip -0.7cm},
\eea
where $(\mu_1,\la_1)$ are free parameters and 
\bea
\nu_1=\frac{\la_1\sqrt{p(p+1)}}{2\sqrt{(p\la_1+1-2q)^2+p\la_1^2+2p\la_1-1}}\,.
\eea
Just as (\ref{Alt_h1}) for $(p,q)=(1,2)$ case, the solution (\ref{Alt_h1Pq}) with nonzero values of $\mu_1$ leads to a wrong asymptotic behavior near $r=r_h$, so it should be discarded.
\item To generalize the limit (\ref{h1asymp0}), we write $h={\tilde h}_1/L^2$ and keep ${\tilde h}_1$ fixed when $L$ goes to infinity in (\ref{SecOrdHpq}). This gives
\bea\label{h1asymp0Pq}
  \frac{qq_-}{2}(q-2)h+qq_-r {\tilde h}_1'+r \left[qs r^2 -\frac{qq_-}{2} {\tilde h}_1\right] \frac{{\tilde h}_1''}{{\tilde h}_1'}=sq^2r^2.
\eea
The solution of this equation is a generalization of (\ref{hRhFlatFin}):
\bea\label{hRhFlatPq}
h=\frac{1}{L^2}\left[\frac{2sr^{q+1}}{(q+1)(r^{q-1}-r^{q-1}_h)}+
\frac{C}{r^{q-1}-r^{q-1}_h}\right],\quad
C=-\frac{2sr_h^{q+1}}{q+1}
\eea 
As in section (\ref{SecSubBlcStr}), the specific value of the integration constant $C$ must be chosen to ensure that function $h$ does not diverge on the horizon. Substitution of the function (\ref{hRhFlatPq}) into (\ref{f123SecOrdPq}) reproduces the well--known metric for black branes with vanishing cosmological constant:
\bea\label{f123flatPq}
f_1=C_1\left[1-\left(\frac{r_h}{r}\right)^{q-1}\right],\quad 
f_3=1-\left(\frac{r_h}{r}\right)^{q-1},\quad f_2=C_2\,.
\eea
\end{enumerate}
\item By solving equation (\ref{SecOrdHpq}) beyond the leading order (\ref{hRhFlatFin}), one can find an expansion in inverse powers of $L$. In the $(p,q)=(1,2)$ case, the expansion is given by (\ref{BeyondLead}), and in general it has a similar form but with more complicated coefficients.
\item In the $r\gg r_h$ region, equation (\ref{SecOrdHpq}) is well approximated by the first order Abel equation (\ref{AbelEqnPq}). In particular, for $r\gg L$, the solution can be expanded as (\ref{hRexpandPq}).
\item Numerical integration of equation (\ref{SecOrdHpq}) with initial conditions (\ref{hRhFlatPq}) near $r=r_h$ would give the unique solution which asymptotes to a generalization of (\ref{hRexpSecOrd}) with specific values of $\alpha$ and $\beta$ for every $r_h$. 
\item Going beyond the $r_h\ll L$ case, one would find the generalizations of (\ref{GenNearHorH}) and (\ref{GenNearHorF3}), which have similar structures, but numerical coefficients depend on $p$ and $q$.
\end{enumerate}

\subsection{Special cases: black holes and maximal branes}
\label{SecSubBlackHoles}

As we have seen in this section, the ansatz (\ref{TheAnsatz}) leads to complicated ODEs which can be analyzed in various asymptotic regions, but don't seem to admit analytic solutions. In this subsection we will focus on two special cases, $p=0$ and $q=1$, which reduce to some well-known geometries.

\bigskip

\noindent
{\bf Black holes: $p=0$}

Setting $p=0$ in (\ref{TheAnsatz}) and substituting the resulting Ricci tensor (\ref{RicciFlat}) into the Einstein's equations (\ref{EinstLam}), we get a system of coupled ODEs for two functions $(f_1,f_3)$. The standard manipulations lead to a well--known solution for black holes in AdS spaces:
\bea\label{BHsolnExpml}
ds^2=-C f dt^2+\frac{dr^2}{f}+r^2d\Omega_q^2,\quad  
f=\frac{r^2}{L^2}+1-\frac{r_h^{q-1}}{r^{q-1}}\,.
\eea
Here $(r_h,C)$ are two integration constants, and the second one can be absorbed in rescaling of time.
Our procedure (\ref{f123SecOrdPq}) for expressing metric components in terms of a single function $h$ fails for this case since there is no function $f_2$. 

\bigskip 

\noindent
{\bf Maximal branes: $q=1$.}

There are two ways to incorporate branes with $q=0$ in the ansatz (\ref{TheAnsatz}): one can either set $q=0$ directly or consider $q=1$ and set $f_2=r^2$. The first option removes the sphere components of the Einstein's equations and re--introduces the freedom in choosing the radial coordinate, which was fixed by the sphere component of the metric (\ref{TheAnsatz}). Therefore the $q=1$ reduction is more direct, and we will follow this path. Starting with the metric
\bea\label{MetrQeq1}
ds^2=-f_1 dt^2+r^2(dx_1^2+\dots dx_p^2)+\frac{1}{f_3} dr^2+r^2 dy^2\,,
\eea
setting $f_2=r^2$ in the Ricci tensor (\ref{RicciFlat}), and solving the resulting Einstein's equations, we find
\bea\label{MaxBranes}
f_1=C_1 f_3,\quad f_2=r^2, \quad f_3=\frac{r^2}{L^2}\left[1-\frac{M}{r^{p+2}}\right]
\eea
Interestingly, the power of $r$ in $f_3$ depends on the number of longitudinal directions along the brane, in contrast to the asymptotically flat case, where only the number of transverse directions matters. As we will see below, the $p$--dependence in (\ref{MaxBranes}) naturally emerges in several examples of non--extremal branes appearing in the AdS/CFT correspondence, and in all these cases such dependence originates from accounting for the directions {\it transverse} to the branes and taking the near--horizon limit.
Before looking at these examples, we note that the Abel equation (\ref{AbelEqnPq}) is fully solvable in the $q=1$ case, and the most general solution for $h(r)$ reads
\bea
r^{4+2p}=\frac{\mu h^{p+2}}{(h-2)^2(hp+4)^{p}}\,.
\eea
Here $\mu$ is a free integration constant.

\bigskip

Let us now look at some well--known examples of non--extremal branes in the near--horizon limit and demonstrate that the solution (\ref{MaxBranes}) is reproduced. Note in all these cases the $p$--dependence in function $f_3$ would arise indirectly, through accounting for the directions {\it transverse} to the branes.

The metric of non--extremal D3-branes in asymptotically--flat space is given by \cite{HorStr}
\bea
ds^2&=&H^{-1/2}\left[-f dt^2+dx_1^2+dx_2^2+dx_3^2\right]+H^{1/2}
\left[\frac{dr^2}{f}+r^2 d\Omega_5^2\right],\\
H&=&1+\frac{R^4}{r^4},\quad f=1-\frac{\rho_0^4}{r^4}\nonumber
\eea
Taking the near--horizon limit, while keeping the ratio $\rho_0/\rho$ fixed, we arrive at a product of a five--dimensional sphere and a metric of the form (\ref{MetrQeq1}) with $p=2$:
\bea
ds^2=-\frac{r^2 f}{R^2} dt^2+\frac{r^2}{R^2}(dx_1^2+dx_2^2+dx_3^2)+\frac{R^2}{r^2 f} dr^2+R^2 d\Omega_5^2\,,
\eea
The metric coefficients match the solution (\ref{MaxBranes}) with $p=2$ after a simple rescaling of the cyclic coordinates.

Next we consider a stack of non--extremal M2 branes:
\bea
ds^2&=&H^{-2/3}\left[-f dt^2+dx_1^2+dx_2^2\right]+H^{1/3}
\left[\frac{d\rho^2}{f}+\rho^2 d\Omega_7^2\right],\\
H&=&1+\frac{R^6}{\rho^6},\quad f=1-\frac{\rho_0^6}{\rho^6}\nonumber
\eea
Taking the near--horizon limit ($\rho\ll R$) and defining a new coordinate $r=\rho^2$, we find the metric of the form (\ref{MetrQeq1}). 
\bea
ds^2=\frac{r^2}{R^4}\left[-f dt^2+dx_1^2+dx_2^2\right]+\frac{R^2 dr^2}{4r^2f}+R^2 d\Omega_7^2,\quad
f=1-\frac{\rho_0^6}{r^3}
\eea
The metric coefficients match the solution (\ref{MaxBranes}) with $p=1$ after a simple rescaling of the cyclic coordinates.

Finally, starting with non--extremal M5 branes,
\bea
ds^2&=&H^{-1/3}\left[-f dt^2+dx_1^2+\dots +dx_5^2\right]+H^{2/3}
\left[\frac{d\rho^2}{f}+\rho^2 d\Omega_4^2\right],\\
H&=&1+\frac{R^3}{\rho^3},\quad f=1-\frac{\rho_0^3}{\rho^3}\nonumber
\eea
and taking a near--horizon limit we arrive at  (\ref{MetrQeq1})--(\ref{MaxBranes}) with $p=4$:
\bea
ds^2=\frac{r^2}{R}\left[-f dt^2+dx_1^2+\dots +dx_5^2\right]+
\frac{4R^2 dr^2}{r^2f}+R^2 d\Omega_4^2,\quad
f=1-\frac{\rho_0^3}{r^6}
\eea
Note that even though in all three examples, the powers in function $f$ originated from {\it transverse} directions in higher dimensions, in the reduced theories with a cosmological constant the powers are determined by the {\it longitudinal} directions, in agreement with (\ref{MaxBranes}).

\section{Black branes with AdS worldvolumes}
\label{SecCurvBrane}

In this section we look for solutions describing black branes without translational symmetry on their worldvolume. The main motivation for studying such objects comes from the well-known solutions for extremal branes in AdS space. While in the absence of cosmological constant, the supersymmetric branes are covered by the harmonic ansatz \cite{HorStr}
\bea
ds^2&=&\frac{1}{f}(-dt^2+dx_1^2+\dots dx_p^2)+f d{\vec y}\cdot d{\vec y},\nn
A_p&=&f^{-1}dt\wedge dx_1\wedge\dots \wedge dx_p, \quad \nabla_y^2 f=0,
\eea
in the presence of an effective cosmological constant the worldvolume of branes becomes curved. For example, fundamental string ending on D3 branes produce curved branes known as bions \cite{Bion}, and in the near--horizon limit such branes have $AdS_2\times S^2$ symmetry \cite{WilsProbe} rather than flat $3+1$--dimensional worldvolume. This symmetry is also present in the gravitational solutions describing supersymmetric D5-D3-F1 configurations in type IIB supergravity, and the metric has the form \cite{WilsSUGRA}:
\bea\label{WilsonMetr}
ds^2=e^{2A} ds_{AdS_2}^2+e^{2B}d\Omega_2^2+e^{2C}d\Omega_4^2+h_{ij}dx^i dx^j\,.
\eea
Although the geometry (\ref{WilsonMetr}) is supported by fluxes rather than a cosmological constant, upon reduction on $S^4$, one gets an effective $\Lambda$ and charged branes with curved worldvolumes. It is natural to expect that some solutions describing curved neutral branes as well, and in this section we will focus on geometries with sphere and AdS factors\footnote{We denoted the metric components by $f_2$ and $f_3$ to stress similarities between (\ref{RicciAdSslice}) and (\ref{RicciFlat}).}:
\bea\label{AdSbrane}
ds^2=f_2 ds_{AdS_p}^2+\frac{dr^2}{f_3}+r^2 d\Omega_q^2,
\eea
where $f_2$ and $f_3$ are functions of one coordinate $r$. 
This metric leads to a diagonal Ricci tensor with components 
\bea\label{RicciAdSslice}
R_{ab}&=&h_{ab}{f_3}\left[-(q-1)-\frac{rf_3'}{2f_3}-\frac{prf_2'}{2f_2}\right]+(q-1)h_{ab},\nn
R_{\alpha\beta}&=&h_{\alpha\beta}\frac{f_2f_3}{4}\left[-\frac{(p-2)f_2'^2}{f_2^2}  - \frac{f_2' f_3'}{f_2f_3}   - \frac{f_2'}{f_2} \frac{2 q}{r} - \frac{2 f_2''}{f_2}\right]-(p-1)h_{\alpha\beta},\\
R_{rr}&=&\frac{1}{4}\left[ - \frac{f_3'}{f_3}(\frac{2 q}{r} 
+ \frac{p f_2'}{f_2}) + \frac{p (f_2'^2 - 2 f_2 f_2'')}{f_2^2}\right],\nonumber
\eea
where $h_{ab}$ and $h_{\alpha\beta}$ are metrics on unit spheres and Anti--de-Sitter spaces. The goal of this section is to explore geometries (\ref{AdSbrane}) which solve the Einstein's equations (\ref{RicciAdSslice}). We will find interesting formal solutions with AdS asymptotics, but unfortunately all resulting geometries contain naked singularities, so the ansatz (\ref{AdSbrane}) does not lead to global physically--interesting solutions. In particular, this implies that to extend the ansatz (\ref{WilsonMetr}) to a non--extremal case, one would have to break the symmetries of the AdS space.

\bigskip

Substituting the metric (\ref{AdSbrane}) and the Ricci tensor (\ref{RicciAdSslice}) into the Einstein's equations (\ref{EinstLam}), one finds an ODE for function $f_3$,
\bea\label{f3eqnAdS}
4 q_-(p + q_-)f_3^2 - 
  2 f_3 \Big[pr^2 f''_3 -f'_3 (p + 2 q_-) r  + 
     2 q_- (p + 2 [q_- - {\bar\Lambda} r^2])\Big]\nn
      + \Big[2 [q_- -  {\bar\Lambda} r^2]- rf_3'\Big]\Big[2 [q_- -  {\bar\Lambda} r^2] -
       (1 + p)rf_3'  \Big]=0,
\eea
as well as the expression for function $f_2$:
\bea\label{f2eqnAdS}
f_2&=&\frac{4 f_3 (p-1) (pr)^2}{4 f_3^2 q_ -(p + q_-) - p_- [
     r f_3'+ 2 {\bar\Lambda} r^2-2 q_-]^2 - 
   4 f_3 F},\\
F&\equiv&-rf_3' (q_- + p) + {\bar\Lambda} (pr)^2 + 
      2 q_- (q_-- {\bar\Lambda} r^2) - 
      p [q_-(q-2)-{\bar\Lambda} r^2(q-3)]. \nonumber
\eea
Recall that ${\bar\La}$ is a conveniently rescaled cosmological constant defined by (\ref{EinstLamDef}) and the dimension of spacetime is $d=p+q+1$. To simplify (\ref{f3eqnAdS}), (\ref{f2eqnAdS}), and subsequent formulas we defined
\bea
q_-\equiv q-1.
\eea
To have interesting spherical and AdS spaces in (\ref{AdSbrane}), we will assume that both $p$ and $q$ are greater than one. Note that the ansatz (\ref{AdSbrane}) with $p=1$ coincides with (\ref{TheAnsatz}) for $p=0$, and the most general solution of the corresponding system is the AdS black hole geometry (\ref{BHsolnExpml}). Let us discuss some properties of equation (\ref{f3eqnAdS}) and its solutions.

The AdS$_p\times$S$^q$ slicing of the AdS space is covered by the ansatz (\ref{AdSbrane}), but it also has a simple relation $f_3=b f_2$ with constant $b$. This reduction in the number of independent functions is analogous to the $f_1=f_2$ constraint (\ref{f123asHpq}) in the case of the flat slicing that reduced the dynamics of flat branes to the first order Abel equation (\ref{AbelEqnPq}). By setting $f_3=b f_2$ in (\ref{RicciAdSslice}), we get much simpler equations that can be easily integrated, and the 
{\it unique} solution reads
\bea\label{f2f3AdSunq}
f_2=\frac{p+q-{\bar\Lambda} r^2}{b(p+q)},\quad f_3=\frac{p+q-{\bar\Lambda} r^2}{p+q},\quad
b=-\frac{{\bar\Lambda}}{p+q}.
\eea
To describe the solution of (\ref{f3eqnAdS}) which approach AdS space at infinity, we should ensure that functions $(f_3,f_2)$ approach (\ref{f2f3AdSunq}) at least in the leading order. Interestingly, there is an exact solution of (\ref{f3eqnAdS})--(\ref{f2eqnAdS}),
\bea\label{f2f3BadExact}
f_3=\frac{q-1}{p+q-1}-\frac{{\bar\Lambda} r^2}{p+q},\quad f_2=-\frac{(p-1)r^2}{q-1}\,,
\eea
that satisfies this condition for $f_3$, but not for $f_2$. Solution (\ref{f2f3BadExact}) leads to a geometry with a wrong signature, so we will not discuss it further.  

Starting with the AdS space (\ref{f2f3AdSunq}), introducing a small perturbation in $f_3$ as
\bea\label{ExpInftyDec20}
f_3=1-\frac{{\bar\Lambda} r^2}{p+q}+\eps g,
\eea
and expanding equation (\ref{f3eqnAdS}) in the first order in parameter $\eps$, one finds a linear second--order ODE for function $g$, and the general solution reads\footnote{Here $F(a,b;c;z]$ is the hypergeometric function.}
\bea\label{ExpInftyDec20a}
g=c_1 r^{1-q}(r^2+L^2)^{-(p-1)/2}+c_2r^2F\left[1,\frac{p+q}{2};\frac{q+3}{2};
-\frac{r^2}{L^2}\right]
\eea
The first term decays at large values of $r$, and the second one goes to constant. Although $c_2$ appears in a subleading order in $f_3$, it modifies the leading behavior of $f_2$:
\bea
f_3=-\frac{{\bar\Lambda} r^2}{p+q}+1+\eps {\tilde c}_2+\dots,\quad
f_2=r^2\left[1-\eps {\tilde c_2}\frac{(p+q-\frac{3}{2})^2-\frac{1}{4}}{p(p-1)}\right]+\dots
\eea
Therefore, for normalizable perturbations one should keep only $c_1$ in (\ref{ExpInftyDec20}), and in the leading order in $\eps$ this gives
\bea\label{tempDec21}
f_3&=&1+\frac{r^2}{L^2}+\eps c_1 r^{1-q}(r^2+L^2)^{-(p-1)/2}+O(\eps^2)\\
f_2&=&L^2\left[1+\frac{r^2}{L^2}+\frac{\eps c_1}{p} r^{1-q}(r^2+L^2)^{-(p-1)/2}+O(\eps^2)\right]
\nonumber
\eea
Using expansion (\ref{tempDec21}) of $f_3$ as an boundary condition at large values of $r$, one can integrate equation (\ref{f3eqnAdS}) to small values of $r$. Numerical simulations show that function $f_3$ always approaches zero at some positive value of $r$, and this value, $r_0$, depends on $\eps c_1$. As in section \ref{SecFlatBranes}, it is instructive to study behavior of functions $(f_2,f_3)$ near $r=r_0$. To avoid unnecessary complications, we will focus on the $(p,q)=(2,2)$ case, and all other geometries (\ref{WilsonMetr}) work in the same way\footnote{Recall that we are assuming that $p,q>1$. The analysis of the $p=1$ and $q=1$ cases is performed separately. }.

\bigskip

\begin{figure}
\begin{tabular}{ccc}
 \includegraphics[width=0.45 \textwidth]{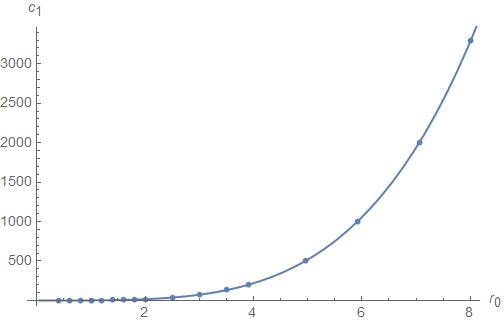}&\ \ &
  \includegraphics[width=0.45 \textwidth]{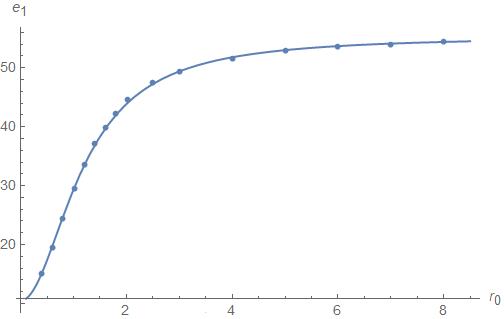}\\
(a)&&(b)
\end{tabular}
\caption{Comparison of approximations (\ref{PolynPade}) with parameters obtained by numerical integration of equation (\ref{tempDec21F3eqn}) for $L=1$:\newline  
(a) ${\tilde c}_1$ as a function of $r_0$;
(b) ${e}_1$ as a function of $r_0$.}
\label{FigOne}
\end{figure}

\begin{figure}
\begin{center}
 \includegraphics[width=0.6 \textwidth]{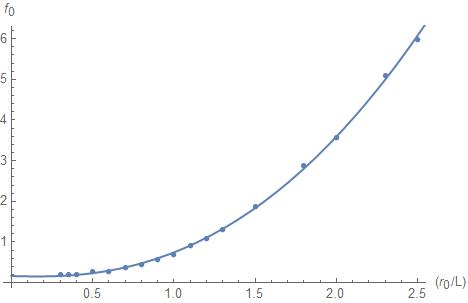}
\end{center}
\caption{Comparison of the approximation (\ref{f0Guess}) with a parameter $f_0$ obtained by numerical integration of equation (\ref{tempDec21F3eqn}) to large negative values of $z$.}
\label{FigTwo}
\end{figure}

For $(p,q)=(2,2)$, expressions (\ref{f3eqnAdS}), (\ref{f2eqnAdS}) simplify: function $f_2$ is given by
\bea
f_2 = \frac{16  r^2 f_3}{12 f_3^2 - 
  f_3 [8 - 12 r f'_3] - [2 -2{\bar\Lambda} r^2 - 
    r f_3']^2},
\eea
and the differential equation for function $f_3$ reads
\bea\label{tempDec21F3eqn}
-12 f_3^2-[rf_3'+2{\bar\Lambda} r^2-2][3rf_3'+2{\bar\Lambda} r^2-2]+
4 f_3[r^2f_3''-2rf_3'-2{\bar\Lambda} r^2+4]=0.
\eea
We are interested in the vicinity of points where function $f_3$ vanishes. Then it is convenient to write 
\bea\label{rR0asZ}
r=r_0+z^2,\quad f_3(r)=z g(z)
\eea
and expand function $g$ in Taylor series in its argument. One finds that the constant term in $g$ must vanish, and the coefficient in front of $z$ can take only one of two values. This leads to two types of solutions:
\begin{enumerate}[(a)]
\item Both $f_3$ an $f_2$ vanish at $z=0$:
\bea
f_3&=&\frac{2(4r_0^2+L^2)}{r_0L^2}z^2+\frac{(4r_0^2-11L^2)}{15r^2_0L^2}z^4+O(z^6)\nn
f_2&=&-\frac{6L^2 r_0}{4r_0^2+L^2}z^2+\frac{9L^2(L^2-4r_0^2)}{5(4r_0^2+L^2)^2}z^4+O(z^6)
\eea
This option leads to a wrong signature through $f_2$, and it cannot be reached by numerical integration starting from (\ref{tempDec21}) large values of $r$. Therefore solutions of this type are irrelevant for analyzing our problem.
\item Only $f_3$ vanishes at $z=0$, while $f_2$ remains finite:
\bea\label{f3f2nearR0}
f_3&=&\frac{4r_0^2+L^2}{r_0L^2}z^2\left[2+\frac{z^2}{r_0}\right]-e_1 z^3+
\frac{15 e_1^2 r_0^3 L^2}{32(4r_0^2+L^2)}z^4+O(z^5)\nn
f_2&=&\frac{128 L^2 r_0^2(4r_0^2+L^2)}{D}+\frac{96 e_1 L^4 r_0^3}{D}z+O(z^2)
\eea
Here
\bea
D=128 (L^2 + 4 r_0^2) (L^2 + 6 r_0^2)-9 e_1^2 r_0^3 L^4
\eea
Numerical integration of equation (\ref{tempDec21F3eqn}) with boundary conditions (\ref{tempDec21}),
\bea\label{tempDec21pq2}
f_3&=&1+\frac{r^2}{L^2}+\frac{{\tilde c}_1}{r^2}+O(r^{-3}),
\eea
indeed gives expansions of this type and determines $(r_0,e_1)$ as functions of ${\tilde c}_1$. In particular, for nontrivial perturbations ${\tilde c}_1$, parameter $e_1$ does not vanish. 
Recalling that 
$dr^2/z^2\simeq 4 dz^2$, we conclude that all components of the metric (\ref{AdSbrane}) remain finite and non--vanishing near $r=r_0$ if one uses the $z$ coordinate. 

Numerical integration of equation (\ref{tempDec21F3eqn}) suggests that 
${\tilde c}_1$ and $e_1$ as functions of $r_0$ are well approximated by ratios of 
polynomials\footnote{For example, asymptotics of ${\tilde c}_1$ at small and large values of $r_0$ suggest that ${\tilde c}_1$ can be approximated by a ratio of polynomials $P_{n+4}(r_0)/P_n(r_0)$. By analyzing such approximations for $n=1,2,3$, we found that the results reduce to (\ref{PolynPade}). Expression for $e_1$ was obtained in the same way.}:
\bea\label{PolynPade}
{\tilde c}_1&=&\frac{0.78 r_0(r_0^2-0.16 r_0 L +1.05 L^2)(r_0^2+0.74 r_0 L +0.67 L^2)}{
L^2(r_0 + 0.6 L)}\,,\nn
e_1&=&
\frac{14.5 L^2+ 55.3 r_0^2}{r_0^{3/2}(1.37 L^2 + r_0^2)}\,.
\eea
Comparison of these approximations with numerical values of 
$\Big({\tilde c}_1(r_0),e_1(r_0)\Big)$ is presented in figure \ref{FigOne}.
\end{enumerate}
Expansions (\ref{f3f2nearR0}) with specific coefficients $(r_0,e_1)$ are obtained by integrating equation (\ref{tempDec21F3eqn}) with boundary conditions (\ref{tempDec21pq2}) and using coordinate transformations (\ref{rR0asZ}) with {\it positive} $z$. However, the resulting space is geodesically incomplte, so after rewriting the metric in $z$ variable, one should continue the geometry to negative values of $z$. This can be done by rewriting the equation (\ref{tempDec21F3eqn}) in terms of the $z$ coordinate and integrating it with boundary conditions (\ref{f3f2nearR0}) at small $z$. Note that due to the presence of odd powers of $z$ in (\ref{f3f2nearR0}) (recall that $e_1\ne 0$), the space at negative values of $z$ looks very different from its $z>0$ counterpart, and numerical analysis shows that it ends in a naked singularity at 
$z=-\infty$. Speficically, numerical integration of equation (\ref{tempDec21F3eqn})  shows that, in the leading order, function $f_3$ grows as
\bea\label{DefineF0}
f_3(z)=f_0\, \left|\frac{z}{L}\right|^{21.9}\left(\frac{r_0}{L}\right)^{12.9}\left[1+O(|z|^{-1})\right],
\eea
where coefficient $f_0$ is well approximated by
\bea\label{f0Guess}
f_0=0.164-0.229(r_0/L)^{1.3}+0.811 (r_0/L)^{2.3}\,.
\eea
Comparison of (\ref{f0Guess}) with numerical results is shown in figure \ref{FigTwo}.
Although we presented the explicit $z$ expansions and numerical results only for geometries (\ref{AdSbrane}) with $(p,q)=(2,2)$, the same analysis is applicable to all $p,q>1$, and it leads to the same qualitative conclusions.    

\bigskip

To summarize, starting with the ansatz (\ref{AdSbrane}) and imposing the asymptotic expansions (\ref{tempDec21}), one obtains a space where radii of AdS and sphere decrease until they reach minimal values at $r=r_0$. The space continues beyond this point via continuation to negative values of the $z$ coordinate. The radii of the AdS space and the sphere start growing again until the space ends with a curvature singularity. Therefore, in the absence of fluxes the ansatz (\ref{AdSbrane}) does not lead to regular geometries. 

\bigskip

To conclude this section, we briefly discuss the special cases $p=1$ and $q=1$. The full solution for the former is given by (\ref{BHsolnExpml}), so we focus on $q=1$. In this case equation (\ref{f3eqnAdS}) simplifies to 
\bea\label{f3eqnAdSS1}
 - 
  2 pf_3 \Big[r^2 f''_3 -rf'_3 
     \Big]
      + \Big[2 {\bar\Lambda} r^2+ rf_3'\Big]\Big[2{\bar\Lambda} r^2+
       (1 + p)rf_3'  \Big]=0.
\eea
Going to variable $r^2$, $f_3(r)=g(r^2)$, we obtain an equation with constant coefficients:
\bea
{\bar\Lambda}^2+{\bar\Lambda}(p+2)g'+(p+1)(g')^2-2pgg''=0.
\eea
One integration of this equation gives a first order ODE
\bea
g=C_1\left[g'+{\bar\Lambda}\right]^2 \left[(p+1)g'+{\bar\Lambda}\right]^{-\frac{2}{p+1}}\,.
\eea
The final solution for $g(r^2)$ can be written in terms of an integral of an inverse function, but the 
answer is not very illuminating. We conclude by observing the for $q=1$, the restriction $f_3=b f_2$ leads to a more general version of (\ref{f2f3AdSunq}),
\bea
q=1:&&f_3=\frac{a {\bar\Lambda}^2}{(p+1)^2}f_2=-\frac{{\bar\Lambda}}{p+1}(r^2+a^2),
\eea
where $a$ is now a free parameter. This extension is not surprising since for $q=1$ the ansatz acquires a freedom of rescaling the radial and ``spherical'' coordinates.

\section{Discussion}
\label{SecDiscus}

In this article we have explored geometries produced by black branes in the presence of a negative cosmological constant. In the past, similar systems have been studied using numerical techniques \cite{CopHor,MannRad}, but by introducing a new parameterization, we were able to get an analytical handle on some properties of black brane geometries in arbitrary dimensions, including their existence, uniqueness, and asymptotic expansions. 

Specifically, we have demonstrated that regular geometries with ${\cal B}=R_t\times S^{q}\times R^{p}$ symmetries and AdS$_{p+q+2}$ asymptotics are governed by one function satisfying the Abel equation (\ref{AbelEqnPq}), which for fixed values of $(p,q)$ looks as simple as (\ref{AbelEqn}). Then the structure of the fixed points (\ref{AttractorInfPq})--(\ref{AttractorZeroPq}) guarantees existence and uniqueness of such regular solutions. Although we have not found explicit solutions for $h$, the Abel equation might provide sufficient information about this function for analyzing excitation of the regular geometry. Due to the symmetry 
${\cal B}$, all perturbations of the background can be decomposed into spherical harmonics on $S^{q}$ and plane waves on $R^{p}$, so the problem reduces to a linear ODE for the radial dependence of excitations, although the coefficients of this equation contain function $h$. It would be interesting to see whether the Abel equation for $h$ 
leads to any simplifications in such coefficients. At the very least, the linear ODE can be analyzed in the regions where $h$ has nice asymptotic expansions found in section \ref{SecSubBrane}. 

Once a horizon is introduced, the Abel equation (\ref{AbelEqnPq}) gets replaced by a more complicated second order ODE (\ref{SecOrdHpq}), but the existence and uniqueness of solutions with regular horizon still persists due to the structure of fixed points and flows between them. The study of excitations is still reduced to linear ODEs, but now the coefficients become more complicated. On the other hand, once the horizon is introduced, one may ask questions about adding charges and rotations to neutral black branes. It would be nice to find some extensions of ``dressing'' techniques, which have been successfully used to construct charged black branes with $\Lambda=0$ \cite{Sen}.

Finally, in section \ref{SecCurvBrane} we explored the ansatz (\ref{AdSbrane}) motivated by extremal branes in AdS space. While we found some interesting local solutions, unfortunately there are no global geometries without naked singularities, so one should look for more general forms of the metric, perhaps by splitting the time coordinate from AdS$_p$. We leave exploration of this possibility for future work.

\section*{Acknowledgements}

This work was supported in part by the DOE grants DE-SC0017962 and
DE-SC0015535. RD also received partial support from the Graduate Research Fellowship at the University at Albany. 

\appendix

\section{Black branes with $\Lambda=0$}
\label{AppFlat}

While this article is dedicated to studying black branes in the presence of a negative cosmological constant, it is instructive to consider solutions of the form (\ref{TheAnsatz}) with $\Lambda=0$. We will do so in this appendix, and find the explicit form of such vacuum solutions. Although we construct large families of local solutions, only the well-known flat $p$--branes give physically interesting geometries.

We consider the metric (\ref{TheAnsatz})
\bea\label{TheAnsatzApp}
ds^2=-f_1 dt^2+f_2(dx_1^2+\dots dx_p^2)+\frac{1}{f_3} dr^2+r^2 d\Omega_q^2
\eea
and impose the vacuum Einstein's equations $R_{\mu\nu}=0$, with Ricci tensor given by (\ref{RicciFlat}). The sphere components of the Einstein's equations become algebraic for $f_3$ if one defines a new function
\bea
f_4={f_1 f_3(f_2)^p}\,.
\eea
Then equations $R_{ab}=0$ give
\bea
\frac{1}{f_3}=1+\frac{1}{q-1}\frac{rf_4'}{2f_4}\,.
\eea
Equations $R_{ij}$ now become
\bea
\frac{2f_2'}{f_2}  - \frac{f_4'}{f_4}  
-\frac{2 q}{r}- \frac{2 f_2''}{f_2'}=0
\eea
To proceed, it is convenient to write function $f_2$ as an exponent, $f_2=e^f$, then the last equation can be integrated as
\bea\label{f4answ}
f_4=\frac{C_1}{r^{2q} [f']^2},
\eea
and the metric takes a very simple form
\bea\label{FlatSolnFinal}
ds^2=-e^{-pf}\frac{(rf')'}{r^{2q} (f')^3} dt^2+e^f(dx_1^2+\dots dx_p^2)-
(q-1)\frac{(rf')'}{f'} dr^2+r^2 d\Omega_q^2
\eea
We used the freedom is rescaling the time coordinate to set $C_1=(q-1)$ in (\ref{f4answ}). At this point the solution (\ref{FlatSolnFinal}) is parameterized by one function $f$, and the remaining 
Einstein's equations, $R_{rr}=R_{tt}=0$ give a fourth order nonlinear ODEs for that function.
Taking a linear combination of these relations, we find a second order equation for $g=f'$
\bea\label{EqnForG}
&&2r^2g(2q+pr g)g''-2r^2(4q+3prg)(g')^2-2p(rg)^2g'\nn
&&\quad-\left[4 q^2 + 4 p q r g + p (1 + p) r^2 g^2\right] (g^2+ 
   r g g')=0
\eea
and this relation along with its integrability condition guarantees that all Einstein's equations are satisfied. Equation (\ref{EqnForG}) is analytically solvable, and to find $g(r)$ one needs to invert
the relation
\bea\label{FlanAnswSoln}
\frac{r^{2q} g^2}{(1+\beta-\alpha rg)(1-\beta-\alpha rg)}
\left[\frac{1+\beta-\alpha rg}{1-\beta+\alpha rg}\right]^{\beta}=C_1
\eea
with two arbitrary integration constants $(\beta,C_1)$. 
Here 
\bea\label{FlanAnswAlpha}
\alpha=-\frac{p}{2q}+\frac{1}{2q}\sqrt{\frac{p}{q-1}}\sqrt{\beta^2q(p+1)-p-q}
\eea
For example, for $p=0$ we get
\bea
g=\frac{g_0}{r^q+r C_2}
\eea
and this leads to the standard black hole geometry. 

Expressions (\ref{FlanAnswSoln}) and (\ref{FlanAnswAlpha}) give the full solution of the problem (\ref{TheAnsatzApp}) unless $f_2$ is a constant function. In the latter case, $f_2=e^f$ is not a good change of variables ($f_1$ and $f_3$ don't have to be constant)s, but the general asymptotically-flat solution is well known:
\bea
f_1=f_3=1-\left[\frac{r_h}{r}\right]^{q-1},\qquad f_2=1.
\eea
It describes black branes. 

To conclude our discussion of the $\Lambda=0$ case, we note that the ansatz (\ref{TheAnsatzApp}) can 
be generalized to 
\bea\
ds^2=-f_1 dt^2+\left[\sum_{k=1}^p h_k dx_k^2\right]+
\frac{1}{f_3} dr^2+r^2 d\Omega_q^2,
\eea
where $(f_1,f_3,h_1,\dots,h_p)$ are functions of the radial coordinate. Then the Einstein's equations 
lead to a generalization of (\ref{FlatSolnFinal}),
\bea
ds^2=e^{-f\sum w_i}\frac{(rf')'}{r^{2q} (f')^3} dt^2+\left[\sum e^{w_if}dx_i^2\right]-\frac{(rf')'}{f'} dr^2+r^2 d\Omega_q^2\,,
\eea
where $w_i$ are arbitrary constants. The solution is fully specified by a set of constants 
${\vec w}=(w_1,\dots w_p)$ and one function $f$. The derivative $g=f'$ still satisfies the transcendental equation (\ref{FlanAnswSoln}). The parameter $\alpha$ depends on the integration constants $(C_1,\beta)$ and vector ${\vec w}$, but the explicit expression is much more complicated than (\ref{FlanAnswAlpha}).

\end{document}